\documentclass[journal,10pt]{IEEEtran}
\usepackage{graphicx}
\usepackage{amsmath}
\usepackage{amssymb}
\usepackage{amsfonts}
\usepackage{cite}
\usepackage{indentfirst}
\usepackage{setspace}
\usepackage{subfig}
\usepackage{float} 
\usepackage{bbm}
\usepackage{dsfont}
\allowdisplaybreaks
\usepackage{lineno}
\usepackage{algorithm} 
\usepackage{algorithmic}
\usepackage{color}

\makeatletter
\newcommand{\removelatexerror}{\let\@latex@error\@gobble}
\makeatother
\ifCLASSINFOpdf
\else
\fi

\begin{document}
	\title{ \LARGE A User Centric Blockage Model for Wireless Networks}
	
	\author{Fran\c{c}ois Baccelli\IEEEauthorrefmark{2},
                Bin Liu\IEEEauthorrefmark{2},
		Laurent Decreusefond,
		and Rongfang Song\vspace{-1em}
	
		\thanks{This work was supported in part by ERC NEMO under the European Union's Horizon 2020 research and innovation programme, grant agreement number 788851 to INRIA, by a Jiangsu Government Scholarship for Overseas Studies under Grant JS-2018-7, Jiangsu Provincial Department of Education, China. (\emph{Corresponding author: Bin Liu})}
		\thanks{\IEEEauthorrefmark{2}Fran\c{c}ois Baccelli and Bin Liu are co-first authors.}
		\thanks{Fran\c{c}ois Baccelli is with Ecole Normale Superieure (ENS) and INRIA in Paris, France (email: Francois.Baccelli@ens.fr).}
		\thanks{Bin Liu, Rongfang Song are with the School of Telecommunication and Information Engineering, Nanjing University of Posts and Telecommunications, China. (e-mail: \{liubin0430, songrf\}@njupt.edu.cn).}
		\thanks{Laurent Decreusefond are with LTCI, Telecom Paris, I.P. Paris, France (e-mail: { laurent.decreusefond}@telecom-paristech.fr).}}

	\markboth{Journal of \LaTeX\ Class Files,~Vol.~XX, No.~X, August~XXXX}%
	{Shell \MakeLowercase{\textit{et al.}}: Bare Demo of IEEEtran.cls for IEEE Journals}
	\maketitle
\begin{abstract}
This paper proposes a cascade blockage model for analyzing the vision that a user has of a wireless network.
This model, inspired by the classical multiplicative cascade models, has a radial structure meant to analyze blockages seen by the receiver at the origin in different angular sectors.
The main novelty is that it is based on the geometry of obstacles and takes the joint blockage phenomenon into account. 
We show on a couple of simple instances that the Laplace transforms of total interference satisfies a functional equation that can be solved efficiently by an iterative scheme. 
This is used to analyze the coverage probability of the receiver and the effect of blockage correlation and penetration loss in both dense and sparse blockage environments. 
Furthermore, this model is used to investigate the effect of blockage correlation on
user beamforming techniques. Another functional equation and its associated iterative algorithm are proposed to derive the coverage performance of the best beam selection in this context.
In addition, the conditional coverage probability is also derived to evaluate the effect of beam switching.
The results not only show that beam selection is quite efficient for multi-beam terminals, but also show how the correlation brought by blockages can be leveraged to accelerate beam 
sweeping and pairing.
\end{abstract}


\IEEEpeerreviewmaketitle

\section{Introduction}
\IEEEPARstart{T}{he} classical distance-based path-loss model with a separate shadowing term has
been employed for decades in wireless network modeling. However, this simple model
falls short in reflecting realistic blockage features where signals from multiple nearby
transmitters are blocked by common obstacles, which leads to the correlated shadowing effect.
Despite significant research efforts in this direction in the recent years, there is still
a clear need for more accurate and yet tractable blockage/shadowing models, particularly
so in the context of 5/6G beamforming.
\subsection{Prior Work}
There is a vast literature on shadowing and the need for correlated shadowing.
Reference \cite{ref1} summarizes the state of the art on correlated shadowing models 
based on parametric path-loss functions of relative/absolute propagation distance and
incidence angle. Reference \cite{ref2} discusses the approach based on
correlated log-normal shadowing random variable. 
All these prior models maintain the analytical tractability at the cost of missing 
the geometric features of blockages. 

Another and more natural way to model correlated shadowing is based on stochastic geometry (SG)
and more precisely on the use of random shape theory to represent the location and the shape of obstacles.
Reference \cite{ref3} surveys the state-of-art on SG based blockage models.  
This is used for both outdoor \cite{ref4} and indoor \cite{ref5} communications.
These random blockage models achieve a good trade-off between tractability and accuracy,
but fail in characterizing spatial correlation.
In \cite{ref6}, a blockage model called the Manhattan-type urban model, leveraged 
Poisson line process to represent blockages. This model was expanded to 3-D indoor
wireless environments \cite{ref7} and outdoor planar networks \cite{ref8}.
These models allow one to take the joint blockage effect into account
but are limited to cities/districts with a Manhattan-type structure.

Beamforming and directional transmission are central in 5G and 6G cellular networks.
This is particularly true for mmWave frequencies where beamforming is
used to compensate for the more severe path loss.
This spurs research efforts on beam scanning, selection, pairing and switch
performance evaluation in different blockage structures. Unfortunately, the vast
majority of these works consider the independent blockage model \cite{ref9}\cite{ref10}
or free space \cite{ref11}\cite{ref12}\cite{ref13}\cite{ref14} for tractability.
They hence fall short in capturing the effect of correlated blockages on beamforming techniques.
There is hence also a need for models allowing one to capture and analyze the effect
of blockage correlation on beamforming techniques.

\subsection{Challenges and Contribution}
The first part of this paper presents a correlated blockage model based on 
a Multiplicative Cascade (MC) model. Such cascades have been successfully used to
describe nonlinear phenomena of multiplicative nature in signal processing \cite{ref16},
network traffic \cite{ref17} etc. for many years. To the best of our knowledge,
this is the first attempt to model correlated blockages in wireless networks using such tools.
This cascade blockage model is parametric and can be used to represent
several types of urban/suburb/rural scenarios, such as city centers, residential districts, 
business centers, suburban areas, etc. It is shown that instead of a representation
in terms of integral forms as in classical SG models, the Laplace transform of interference 
can be obtained by iterative algorithms derived from the cascade functional equation. 
This is then used to derive the downlink coverage probability. 

The second part of the paper is focused on the impact of blockage correlation on beamforming.
The cascade is again used to establish a functional equation for the joint
Laplace transform of interference in all angular sectors. This again leads
to iterative algorithms which can are used to obtain the coverage performance
of the best beam selection scheme. We also analyze the conditional coverage probability
in case of a beam switch. We illustrate the practical use of this in the
context of beam scanning and pairing.

The main contributions of this paper can be summarized as follows: 

\emph{1)} A basic cascade model and two extensions are proposed to describe blockage environments.
These models capture blockage correlation effects. Compared with independent models used
in prior work, our models are equally tractable and more realistic. They also
show that the independent blockage models underestimate the overall system performance. 

\emph{2)} For omnidirectional User Equipment (UE), new functional equations and
iterative algorithms are proposed
to derive Laplace transforms of total interference.

\emph{3)} A beamforming capable UE can sweep the beamforming directions and choose the beam with the
maximal Signal to Interference Ratio (SIR) as its serving beam. The blockage correlation
complicates the computation of the best beam. We propose another functional equation
and its associated iterative algorithm. This allows one to obtain the coverage probability
in this case. 

\emph{4)} We leverage the cascade model to derive the conditional coverage probability
after a beam switch. The analysis shows how to scan beams from a reference beam
in order to shorten the beam scanning duration.

\emph{5)} Simulation results validate the accuracy of analysis and are used to gain further
insights on the impact of blockage correlation on coverage performance for both
omnidirectional and beamforming UEs.  

\subsection{Organization and Notation}
The rest of paper is organized as follows. Section II introduces the system model.
In Section III, we propose the iterative algorithm allowing one to evaluate coverage performance 
for omnidirectional UEs. Two variants of the basic cascade model of Section II 
allowing one to represent other blockage environments are discussed in Section IV. 
In Section V, we introduce the problem of the best beam selection
and beam switch in correlated blockages. This is based on an analysis of the joint distribution
of interference in different angular sectors. An algorithm to compute this joint distribution 
is also give. This is the basis of the evaluation of coverage probability of the best beam policy
when users have beamforming functionalities.
Beam switching performance is further evaluated in Section VI.
Finally, we conclude the paper in Section VII. Table I gathers the notation used throughout.

\section{System Model}
In a cellular network, the transmission between BSs and users are often blocked by
obstacles like buildings, cars and walls. In most cities, towns or districts, blockages
have the following basic physical features: \emph{Firstly,} their locations are 
often aligned along with certain geometric objects such as avenues, streets, roads, and
rivers thanks to urban planning. \emph{Secondly}, most blockages are not as regular
as those in Manhattan-type cities. 
\emph{Thirdly,} the nearer the blockage, the larger the angle with which a given user sees it.

Consider BSs deployed uniformly on the plane and a typical user located at the origin.
Assume obstacles hindering electromagnetic propagation to be also deployed in the plane.
These obstacles can have very different structures depending on the environment 
(urban/suburb/rural areas). For example, buildings are organized in dense and
locally periodic structures in urban areas, whereas they may have a sparse and random 
structure in rural areas. These features are partially captured by the following radial model.

\begin{figure}[!t]
	\centering
	\includegraphics[width=2.8in]{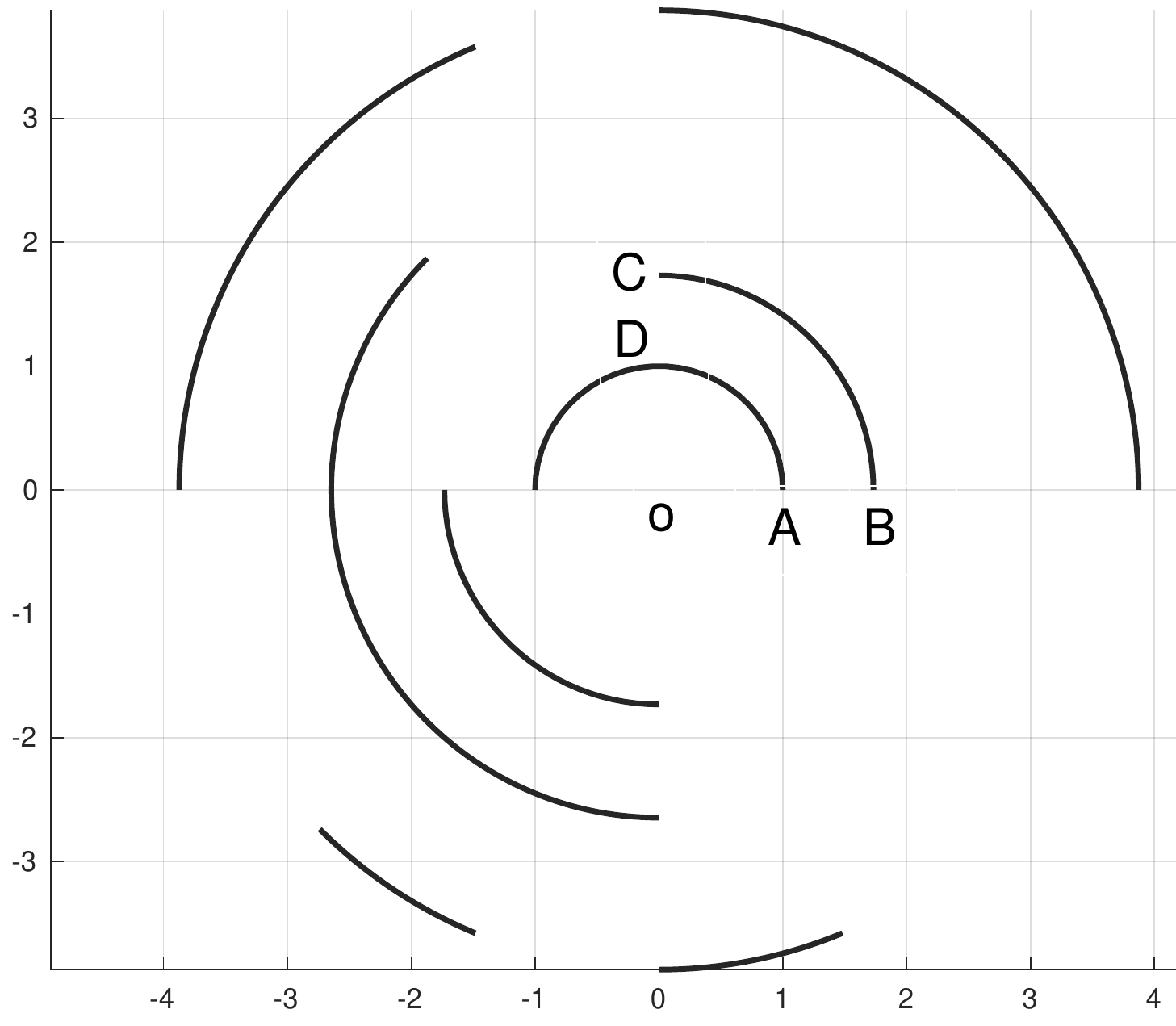}
	\captionsetup{singlelinecheck=off}
	\hfil
	\caption{An example of basic cascade blockage model. The origin represents typical user and the solid curves are blockages.}
	\label{fig1}
	\vspace{-0.25cm} 
\end{figure}

\subsection{The Basic Radial Blockage Model}
A cascade is an iterative procedure that divides a given set into smaller and smaller
sets using some subdivision rules \cite{ref16}\cite{ref17}. In the proposed model,
the blockages are created according to a cascade procedure operated on some angular domain,
e.g., $[0,2\pi]$ if the antenna of the receiver is omnidirectional. In the proposed model,
the blockages are arranged as a collection of circle arcs. The circles containing these arcs
are all centered at the origin and are numbered from 1 to $\infty$. The $n$-th circle has a
radius $R_n$. We assume $0<R_1<R_2<\cdots$ to take the fact that there are nearby and
more distant obstacles into account. In the basic model described here, at the first cascade 
iteration (stage number $n=1$), if some first stage blockages exist, they occupy half the angular 
domain, e.g., $[0,\pi]$ or $[\pi, 2\pi]$ on the circle with radius $R_1$. At stage $n=2$,
each potential blockage interval of stage $n=1$ is divided into two equal length subintervals, 
and as for stage 1, obstacles of stage 2 can be present and then occupy some of these subintervals, i.e.,
$[0,\pi/2],[\pi/2,\pi],[\pi, 3\pi/2]$, and $[3\pi/2,2\pi]$. This goes on with stage 3, and so on.
An instance of such a random collection of obstacles is depicted in Fig.\ref{fig1}. This leads to an organization 
of obstacles as an infinite binary tree, with the obstacles
of stage $n$ being located on the $n$-th circle. Except for the root, each node in
this tree is either unblocked (absence) or blocked (presence of an obstacle on the arc in question).

Whether one of the subintervals of the $n$-th stage is blocked or not depends on the scenario.
A simple model is that where each subinterval is declared blocked randomly and independently with some
given probability capturing the angular density of obstacles. Another natural model is that
where obstacles are displayed in a periodic manner on each circle. There is a rich variety
of such models beyond this first dichotomy (random versus periodic). One can vary the sequence of radii.
One can also consider other schemes than this binary one.

In the sequel, by a \emph{box}, we mean  
any annular region delimited by two adjacent
circles and two potential blockage arcs on these circles.
An instance of such a
box is the annular region ABCD in Fig.\ref{fig1}. 

Below we first focus on the basic model with this binary subdivision for simplicity.
In order to make the analysis easier, we will also assume the $R_n$ are chosen in such a way
that the area of each box is equal for each stage.
This requires that 
\begin{equation}
	R_{n+1}^2=R_n^2+ 2^n R^2,\quad n\ge 1,
\end{equation}
with $R_1=R$, an arbitrarily
chosen positive constant. Hence, the radius of $n$-th stage is set to $R_n=R\sqrt{2^n-1}$.
In particular, we set $R_0=0$.

\begin{table}[!t]
	\centering
	\caption{System Model Parameters}\label{tab:tab2}
	\begin{tabular}{|c|c|}
		\hline
		Notation & Description \\\hline
		${\Phi}$, ${\lambda}$ & BSs PPP with intensity ${\lambda}$\\
		\hline
		$R, R_n$&network parameter and radius of n-th stage\\
		\hline
		$p,q$ & blockage prob. and non-blockage prob.\\
		\hline
		$N$&max. stage number\\
		\hline
		$h_x,h^i$&Rayleigh fading of BS $x$ and i-th beam\\
		\hline
		$K$& blockage penetration loss factor, $K\le 1$\\
		\hline
		$I_n $&total interf. in n-th stage  \\
		\hline
		$I_{n,sind} $&total interf. of semi-indep. model in n-th stage  \\
		\hline
		$I_{n,rep} $&total interf. of repulsive model in n-th stage  \\
		\hline
		$I_n^l$&total interf. in n-th stage inside beam $l$ \\
		\hline
		$A_n(V)$&local interf. in n-th stage with vol. V\\
		\hline
		${\cal A}_n(s,V)$ &LT of local interf. in n-th stage with vol. V\\
		\hline
		$k$& $2^k$ is beam number or UE antenna number\\
		\hline
	\end{tabular}
\end{table}

\subsection{Network Model}
Consider the downlink of an interference-limited cellular network. In this blockage environment,
BSs are assumed to be deployed as a homogeneous Poisson Point Process (PPP) ${\Phi}$ with intensity
${\lambda}$ in ${\mathbb{R}^2}$. Each BS is assumed to have unit transmission power.
The typical user is assumed to be located at the origin. The idea being that the
radial structure that it sees is a snapshot of its current obstacle environment.
Its terminal is first assumed here to have an omnidirectional reception antenna.
The case of a terminal with several panels and beamforming functionalities will be discussed
in the forthcoming sections.
Each link between a BS and the typical user suffers from independent and identically
distributed (i.i.d.) Rayleigh fades $h_x$, (${x \in \Phi } $), namely the signal power
is multiplied by a random variable with an $Exp(1)$ distribution. The large scale path-loss
is neglected throughout. This simplification is justified by the fact that
the loss caused by the blockages dominates.

All blockages are assumed to have the same penetration loss denoted by $K, (K\in [0,1))$.
Let us stress that there is no problem extending the framework described below to the case where this constant
is replaced by an independent random variable with support in $[0,1]$.
Under the above assumptions, all boxes at all stages have the same volume
$V=\pi R^2/2$. In the basic independent model, each subinterval is independently blocked
with probability $p, (p\in (0, 1])$ in case of a finite stage blockage environment.
In case of an infinite stage blockage environment, we assume that $p>\frac 1 2$ in order to 
guarantee the finiteness (and integrability of the Light-Of-Sight (LOS) region).

Since blockages are symmetric in law with respect to the $x$ axis, the analysis will
focus on the blockages on the upper half of the plane, as shown in Fig.\ref{fig2}.

\section{Interference and Coverage}
\subsection{Distribution of Total Interference}
The total interference seen by the typical user is
\begin{equation}
	{J} = \sum _{x \in \Phi }^{}{h_x}{K^{{N_x}}},
\end{equation}
where $N_x$ denotes the total number of blockages between the origin and the BS at position $x$.
\begin{figure}[!t]
	\centering
	\includegraphics[width=3in]{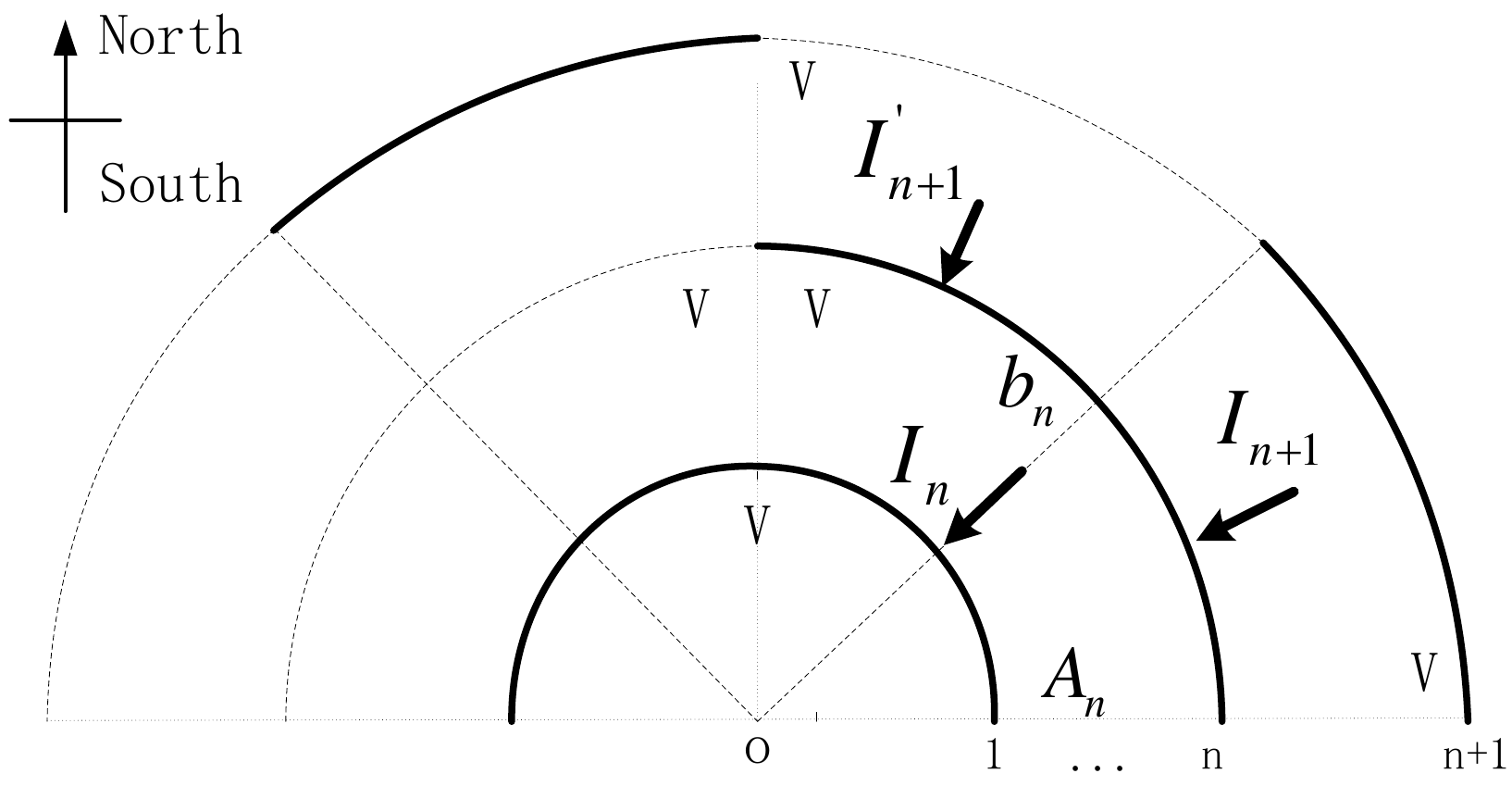}
	\captionsetup{singlelinecheck=off}
	\hfil
	\caption{An illustration of interference relation between adjacent stages.}
	\label{fig2}
	\vspace{-0.25cm} 
\end{figure}

Let $A(V)$ be the interference created by (and in) a box of volume $V$. The Laplace transform of $A(V)$ is
\begin{align}
	{{{\cal A}(s,V)}} :=& \mathbb{E}\left[ e^{-sA(V)} \right] ={\sum\nolimits_{k = 0}^\infty  {(\lambda V)} ^k}{e^{ - \lambda V}}{F^k}(s)/k! \notag\\
	=& \exp ( - \lambda V(1 - F(s))),
\end{align}
where $F(s)$ is the Laplace transform of Rayleigh fading with parameter one, i.e., $F(s)=1/(1+s)$. 

Consider a box of the $n$-th stage, namely between circles $n-1$ and $n$, $n\ge 1$,
and with angular stretch $[\theta,\theta + \frac{2 \pi}{2^{n}}]$. 
Let $I_n$ be the total interference seen in this box and stemming from the
region $\mathbb R^2\setminus B(0,R_{n-1}) \cap C(0,[\theta,\theta + \frac{2 \pi}{2^{n}}])$,
with $B(0,r)$ the ball of center 0 and radius $r$ and $C(0,[\phi,\psi])$ the cone of apex 0
and angle $[\phi,\psi]$. In other words, $I_n$ consists of the sum of two terms:
(a) the interference coming incurred in this box from the regions at distance more than
$R_{n}$ from the origin and in the angular interval $[\theta,\theta + \frac{2 \pi}{2^{n}}])$,
and \emph{(b)} the interference $A_n(V)$ created and incurred in this very box, as shown in Fig.\ref{fig2}.
It should be clear that for all such boxes, $I_n$ has the same distributions as $I_1$
(this follows from the fact that each is built in the same stochastic way from a binary tree and from
the fact that all boxes have the same volume).   
Hence, in particular,
\begin{align}
{I_1} & = (I_{2}+ I_{2}^{'}){K^{b_{1}}} + {A_1(V)},
\end{align}
where the random variables $(I_{2},I_{2},A_1(V))$ are independent and $I_1,I_2$, and $I_2'$ have the same 
distribution. In addition, $b_1 \in \{ 0,1\}$ is a Bernoulli random variable which 
equals $1$ with probability $p$ and $0$ with probability $q=1-p$, and is independent of {{$(I_2,I_2',A_1(V))$.}}
Hence, the Laplace transform of $I:=I_1$ satisfies the functional equation
\begin{align}
\label{Lap}
	{{\cal L}_{{I}}}(s)  &= \mathbb{E}[{e^{ - sI_1}}] \notag\\
	&=\mathbb{E}[{e^{ - s{I_{2}}{K^{b_1}} - sI_{2}^{'}{K^{b_1}}  }}]\mathbb{E}[{e^{ - s{A_1(V)}}}]\notag\\
	&= \left(p\mathbb{E}[{e^{ - Ks{I_{2}} - KsI_{2}^{'} }}] + q\mathbb{E}[{e^{ - s{I_{2}}- s{I_{2}^{'}}   }}] \right){{\cal A}(s,V)} \notag\notag\\
	&=\left( p {{\cal L}_{{I}}}(Ks)^2 +q{{\cal L}_{{I}}}(s)^2 \right){{\cal A}(s,V)} .
\end{align}

The total interference received at the typical user is
${{\cal L}_{J}(s)={\cal L}_{I}(s)}^2$ due to the symmetric blockage and
BS distributions in the northern and southern ${\mathbb{R}^2}$ half-planes, plus independence.

{{Here is a natural iterative scheme for solving the last functional equation:
$Q^{(0)}(s)={\cal A}(s,V)$ and for all $n\ge 0$,
\begin{align}
	Q^{(n+1)}(s)  =  \left( p Q^{(n)}(Ks)^2 +qQ^{(n)}(s)^2 \right){{\cal A}(s,V)}.
\end{align}
The distribution of $I_1$ is that with Laplace transform $Q^{(\infty)}(s)$.

It is easy to extend this approach to a finite domain (cascade) case. Assume there are only $N$ circles.
Then the Laplace transform $M^{(N)}(s)$ of $I_N$ is equal to ${\cal A}(s,V)$ and
more generally, the Laplace transform $M_n$ of $I_n$, $1\le n< N$ is obtained by 
\begin{align}
\label{Lapbis}
	M^{(n)}(s)  =  \left( p M^{(n+1)}(Ks)^2 +qM^{(n+1)}(s)^2 \right){{\cal A}(s,V)}.
\end{align}
Note again that there is no need to assume that $p>1/2$ in the finite service area case.
This is only needed in the infinite case to guarantee that the total LOS interference be finite.
}}

Note that the complexity (number
of operations) of this algorithm is linear in $N$. For the infinite domain case,
an approximation of depth $n$ has a complexity which is linear in $n$.

Ler $G$ and $h$ respectively denote the antenna gain and the Rayleigh fading of parameter 1
of the serving BS. Since we assume unit transmission power, the probability of $\theta$-coverage
(defined as the probability that the SIR exceeds $\theta$) by this BS is 
\begin{equation}
P_{{\rm{cov}}}(\theta )= \mathbb{P}\{ \frac{{Gh}}{I} \ge \theta \}  = \mathbb{E}[{e^{ - \theta I/G}}]={\cal L}_I\left( \frac \theta G\right).
\end{equation}

\subsection{Simulation Results}
\begin{figure}[!t]
		\centering
	\subfloat[$K$ variations]{\includegraphics[width=2.8in]{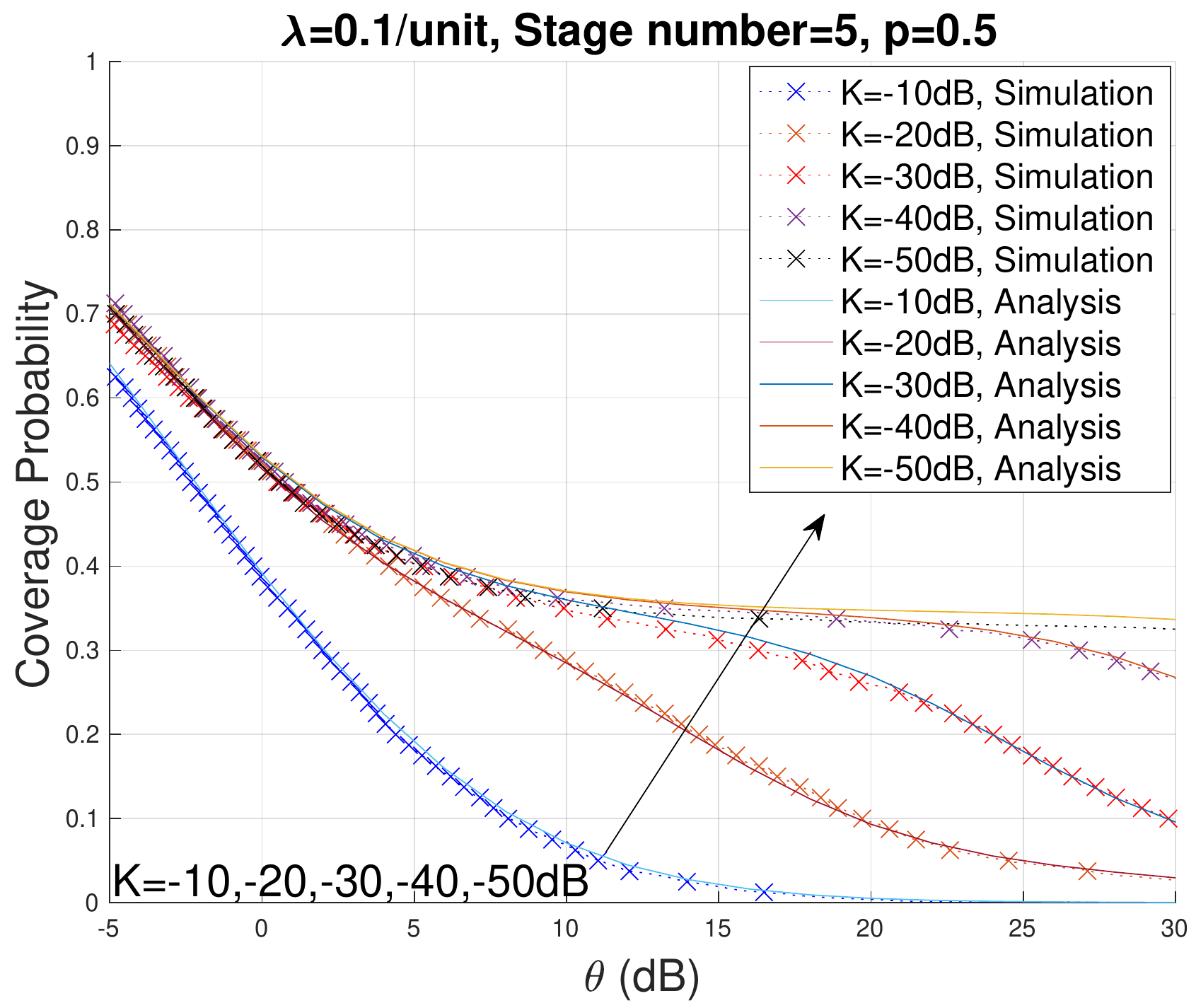}
		\label{fig31}}
	
	\subfloat[Sparse/moderate/dense networks]{\includegraphics[width=2.8in]{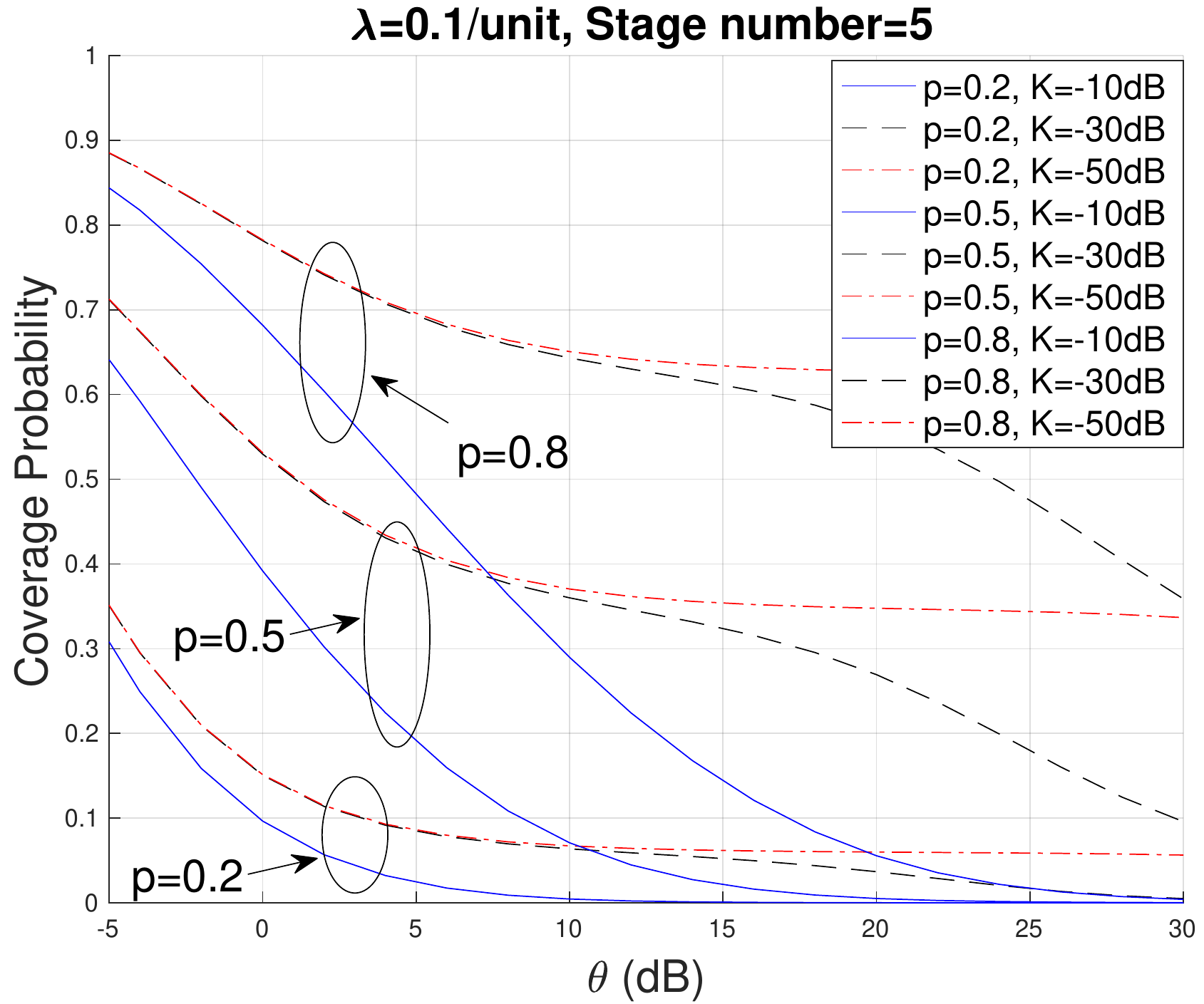}
		\label{fig32}}
	\captionsetup{singlelinecheck=off}
	\hfil
	\caption{Coverage performance comparisons in the basic model ($\lambda=0.1/unit$, max. stage no. $N=5$.)}
	\vspace{-0.25cm} 
\end{figure}

We now assume that a virtual LOS serving BS of the typical user is added to $\Phi$,
which only suffers from Rayleigh fading. The knowledge of the distribution of interference
allows one to get a simple formula for the probability of coverage for a threshold $\theta$,
which is the probability that the SIR exceeds $\theta$.

In Fig.\ref{fig31}, we compare the coverage probabilities of different penetration losses under 
the virtual BS association in an interference-limited network. Obviously, a higher penetration
loss reduces Non-LOS (NLOS) interference from boxes and gives rise to a higher coverage probability.
When $K<-40dB$, the interference variations due to different blockage penetration losses is
negligible for $\theta<25dB$.

Fig.\ref{fig32} compares the coverage probabilities of sparse ($p=0.2$), moderate ($p=0.5$) and dense 
($p=0.8$) blockage scenarios when the penetration loss $K$ varies from -10dB to -50dB.
As expected, the coverage probability gaps between different $K$ cases are very small
for sparse blockage scenarios, but not negligible for dense ones.
This is in line with the intuition that fewer blockages lead to smaller variations.
In addition, the sparse blockage scenario has lower coverage probabilities than the other
scenarios, because fewer blockages make sparse scenarios more vulnerable to LOS interference
and lead to lower SIR. Finally, the dense blockage scenario is more sensitive to the
threshold $\theta$ due to the higher SIR induced by more blockages. 

\section{Variants of the Basic Cascade Model}
\subsection{A Less Correlated Cascade Model}
The basic cascade model can be adapted to represent a less correlated blockage scenario.
We construct a new cascade model by independently setting each half of an interval of the
basic model to be a blockage or non-blockage state. 
This variant corresponds to scenarios
where blockage sizes are smaller than those in the basic model and it hence exhibits more
randomness, since it has more blockage location freedom in each stage.
We denote the interference observed by the typical user in this 
less correlated model by ${\cal L}_{J,lc}$.
In this model, the volume $V$ that was
used in the basic model has to be replaced by $\tilde V \buildrel \Delta \over =V/2$. 
Let ${\cal L}_{\tilde I_1}$ be the interference 
at the origin due to the quarter plane $[0,\pi/2]$. We have
${\cal L}_{J,lc}= {\cal L}_{\tilde I_1}^4$.
Let $\tilde I_n$ denote the 
total interference seen in the box in the $n$-th stage  and stemming from the
region $\mathbb R^2\setminus B(0,R_{n-1}) \cap C(0,[\theta,\theta + \frac{2 \pi}{2^{n+1}}])$.
By the same arguments as above
\begin{align}
{\tilde I_{n}} &= {\tilde I_{n + 1}}{K^{{b_{n}}}} + \tilde I_{n + 1}^{'}{K^{b_{n}}} + {A_n(\tilde V)},
\end{align}
where $b_n\in \{ 0,1\}, (\forall n \in \mathbb{N} ) $ represent the presence of blockage in the subinterval. Then, by the same arguments as above,
\begin{equation}
{{\cal L}_{{\tilde I_{n}}}}(s)
 =\left( p{{\cal L}_{{\tilde I_{n + 1}}}}(Ks)^2 +q{{\cal L}_{{\tilde I_{n + 1}}}}(s)^2 \right)
{{\cal{A}}(s,\tilde V)}.
\end{equation}

\textbf{Lemma 1:} The basic cascade model has larger Laplace transform of total interference,
thereby higher coverage probability, than the less correlated model.

\emph{Proof}: 
We first prove this by for the model with $N$ stages. We have 
\begin{align*}
{{\cal L}_{{\tilde I_{N}}}}(s)^2 = 
{{\cal L}_{{I_{N}}}}(s)&={{\cal{A}}(s,\tilde V)}^2={{\cal{A}}(s,V)}.
\end{align*}
Take as induction assumption that 
$${{\cal L}_{{\tilde I_{N-k}}}}(s)^2 \le 
{{\cal L}_{{I_{N-k}}}}(s).$$
Then
\begin{eqnarray*}
{{\cal L}_{{\tilde I_{N-k-1}}}}(s)^2 & = & 
 \left( p{{\cal L}_{{\tilde I_{N-k}}}}(Ks)^2 +q{{\cal L}_{{\tilde I_{N-k}}}}(s)^2 \right)^2
{{\cal{A}}(s,\tilde V)}^2 \\
& = &\left( p{{\cal L}_{{\tilde I_{N-k}}}}(Ks)^2 +q{{\cal L}_{{\tilde I_{N-k}}}}(s)^2 \right)^2
{{\cal{A}}(s, V)} \\
& \le & \left( p{{\cal L}_{{I_{N-k}}}}(Ks) +q{{\cal L}_{{I_{N-k}}}}(s) \right)^2
{{\cal{A}}(s, V)} \\
&  \le & \left( p{{\cal L}_{{I_{N-k}}}}(Ks)^2 +q{{\cal L}_{{I_{N-k}}}}(s)^2 \right)
{{\cal{A}}(s, V)}\\
& = &
{{\cal L}_{{I_{N-k-1}}}}(s),
\end{eqnarray*}
where the first inequality is due to the induction assumption and the second to convexity.

{{The result for the infinite cascade model is then obtained when letting $N$ to infinity.}}

$\hfill\Box$

\begin{figure}[!t]
	\centering
	\subfloat[Basic vs. Less correlated Models]{\includegraphics[width=2.8in]{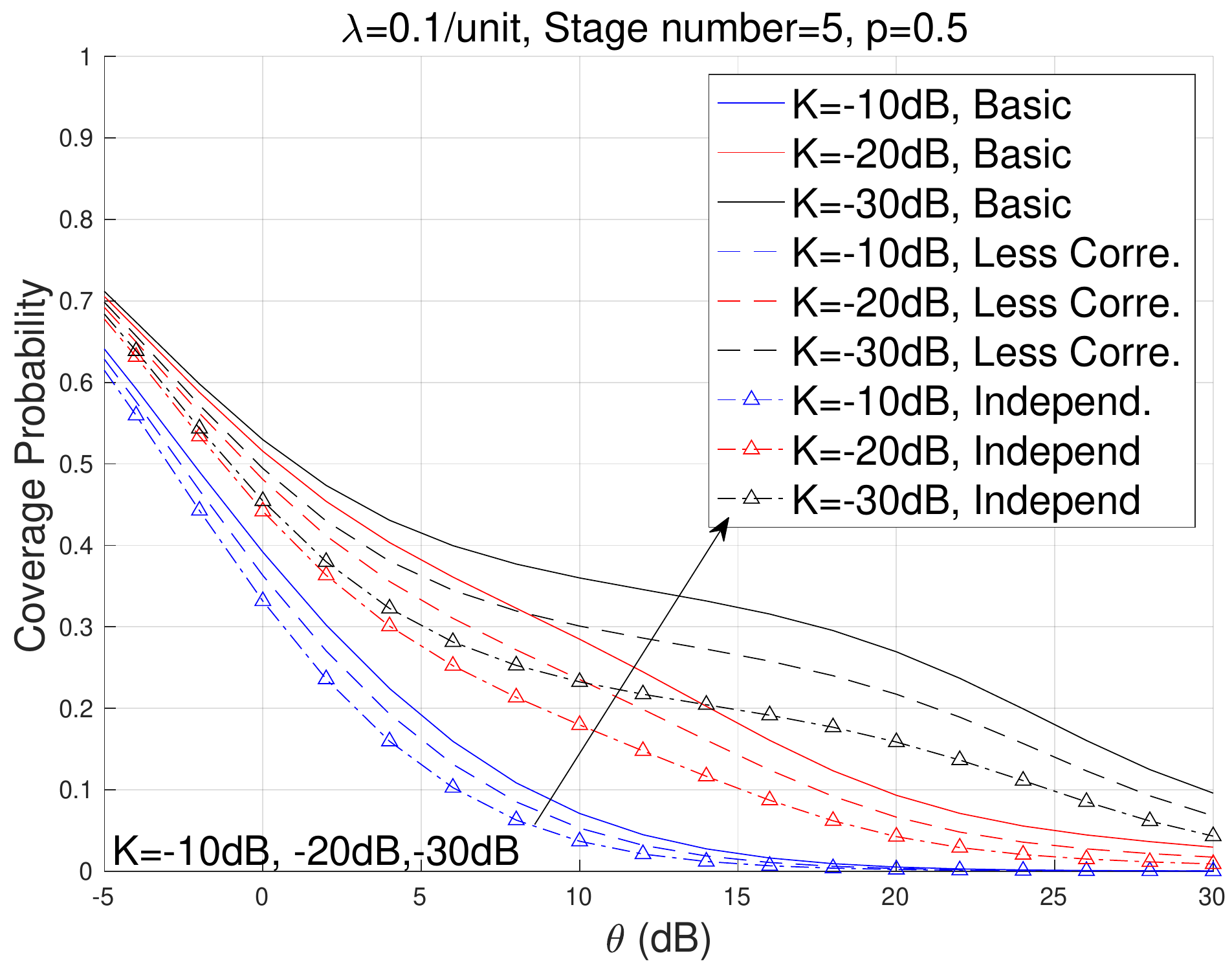}
		\label{fig41}}
	
	\subfloat[Basic vs. Periodic Models]{\includegraphics[width=2.8in]{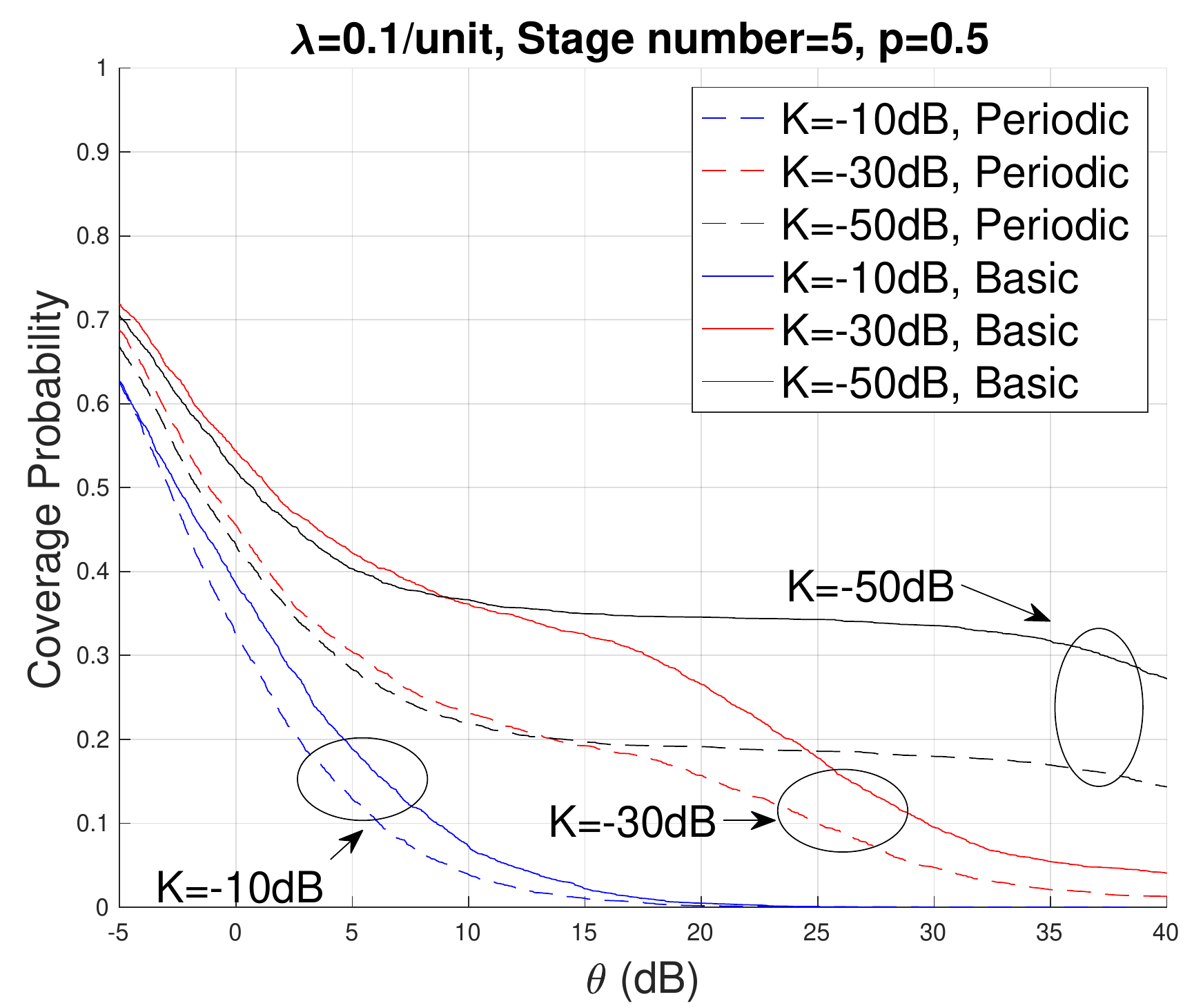}
		\label{fig42}}
	\captionsetup{singlelinecheck=off}
	\hfil
	\caption{Coverage performance comparisons ($\lambda=0.1/unit$, max. stage no. $N=5$)}
	\vspace{-0.25cm} 
\end{figure}

In Fig.\ref{fig41}, the coverage probabilities of the two cascade models are plotted when
the virtual BS model is assumed. The basic model has higher performance gains under all kinds of
penetration losses, especially when $K$ is very small. It also implies that the Shannon 
capacity of the basic model, which is a monotonic increasing function of coverage probability,
is higher than that of the less correlated model. This is yet another illustration
of the principle stated in [8] that the positive correlation created by obstacle shadowing
is {\em beneficial} to wireless communications.

\subsection{Periodic Cascade Model}

The basic model can be revised to emulate well-planned areas such as residential areas/business centers.
One of the two subintervals is declared as a blockage randomly and the next subinterval is designated 
as the non-blockage. This leads to a periodic model with a probability for an interval to be blocked 
equal to 1/2. This model is only studied here in the finite $N$ case.

The iterative algorithm for the Laplace transform of interference at the n-th stage $I_{n,per}$ is
\begin{align}
	{{\cal L}_{{I_{n,per},}}}(s) &= \mathbb{E}[{e^{ - s{I_{n,per}}}}] \notag\\
	&= {{\cal L}_{{I_{n + 1,per}}}}(Ks){{\cal L}_{{I_{n + 1,per}}}}(s){\cal{A}}(s,V).
\end{align}

Fig.\ref{fig42} shows the coverage performance of the two models. The basic model has significant
coverage performance gains compared to the periodic one,
especially so at high penetration losses. 

{{
\subsection{Independent Model}
The independent blockage model \cite{ref4} is that where each link between
the typical user and a BS suffers from an {\em independent} blockage penetration loss with a distribution that
depends on the distance between them. In the present situation, this distribution can be evaluated as follows.
For a BS at distance $r$ from the origin, let $n(r)$ be the integer such that $n(r)\le r <n(r)+1$.
Using the fact that there are $n(r)$ potential obstacles between the BS and the origin,
which are independently open or closed,
we get that the blockage penetration loss $S(r)$ is a random variable with support
on $\{1,K,K^2,\ldots,K^{n(r)}\}$ and such that
\begin{align}
\mathbb{P}[S(r)=K^l] ={n(r) \choose l}p^{l}q^{n(r)-l},\quad l=0,\ldots n(r).
\end{align}

In this independent model, the interference at the origin is the Poisson shot-noise
$$I_{\mathrm{ind}} = \sum_{X_i\in \Phi} h_i S_i.$$
Here $\Phi$ is the BS Poisson PP, the random variables $h_i$ are independent Rayleigh fades
with mean 1, whereas the random variables $S_i$ is conditionally independent (given $\Phi$ and $h=\{h_i\}$),
and such that $S_i$ is distributed like $S(r_i)$ with $r_i=||X_i||$.
By standard arguments,
\begin{eqnarray}
& &\hspace{-.5cm} \mathbb{E} [ \exp(-s I_{\mathrm{ind}})] =  \\
& & \exp\left( -\lambda 2\pi \int_0^\infty \left(1- \mathbb{E} \left[ \frac{1}{1+sS(r)}\right] \right) rdr
\right), \nonumber
\end{eqnarray}
with $S(r)$ the random variable defined above.
Hence
\begin{eqnarray}
& &\hspace{-.7cm} \mathbb{E} [ \exp(-s I_{\mathrm{ind}})] =  \\
& & \hspace{-.7cm}\exp
\left(
-\lambda 2\pi
\int_0^\infty \left(1- \sum_{l=0}^{n(r)} {n(r) \choose l}p^{l}q^{n(r)-l}
\frac{1}{1+sK^l} \right) rdr \right).\nonumber 
\end{eqnarray}

The probability of coverage under this independent model is then easily deduced from this expression.
}}

\section{Beamforming in UE - Best Beam Selection}
\subsection{Motivation}
{{Up to this point in the paper, it was assumed that all UEs are omnidirectional.
This section and the next are focused on the case where UEs are equipped with one or more panels,
and where each panel has beamforming functionalities.
Such a scenario will become a reality in a few years, particularly so in the millimeter wave
(mmWave) case. In this setting, panel switch and beam selection/switch will become a particularly
important matter as this will allow the network to
compensate for the severe path loss in these bands.
However, this will require both sides to perform
exhaustive sweeping through all possible beamforming directions until the UE steers its
beam bore-sight toward the best serving BS's transmission beam.}}
This search procedure is a part of initial access
\cite{ref9}\cite{ref10}\cite{ref11}\cite{ref12}\cite{ref13}\cite{ref14}\cite{ref15}.
This beam pairing procedure will be re-initiated in case of severe blockage or UE/BS mobility.
It can be refined by hierarchical search (combining both coarse-grained sector and beam refinement phases)
or omni-directionally reception \cite{ref10} to shorten the association procedure duration.   

There is a large corpus of prior work on the combination of these network paradigms with beamforming
in different blockage scenarios and for different link techniques. Unfortunately, none of these
capture the impact of correlated blockage on beam selection and beam switching.

Here are interesting and to the best of our knowledge unresolved questions:
\emph{(a)} What is the effect of blockage/shadowing correlation on beam selection?
\emph{(b)} What is the coverage probability under the best beam or the best beam pair policy?
\emph{(c)} {{What are the performance improvements after a beam switch caused by some
blockage event in such a correlated blockage environment?}}

In the following, we leverage our cascade blockage model to study these questions.
\begin{figure}[!t]
	\centering
	\includegraphics[width=3in]{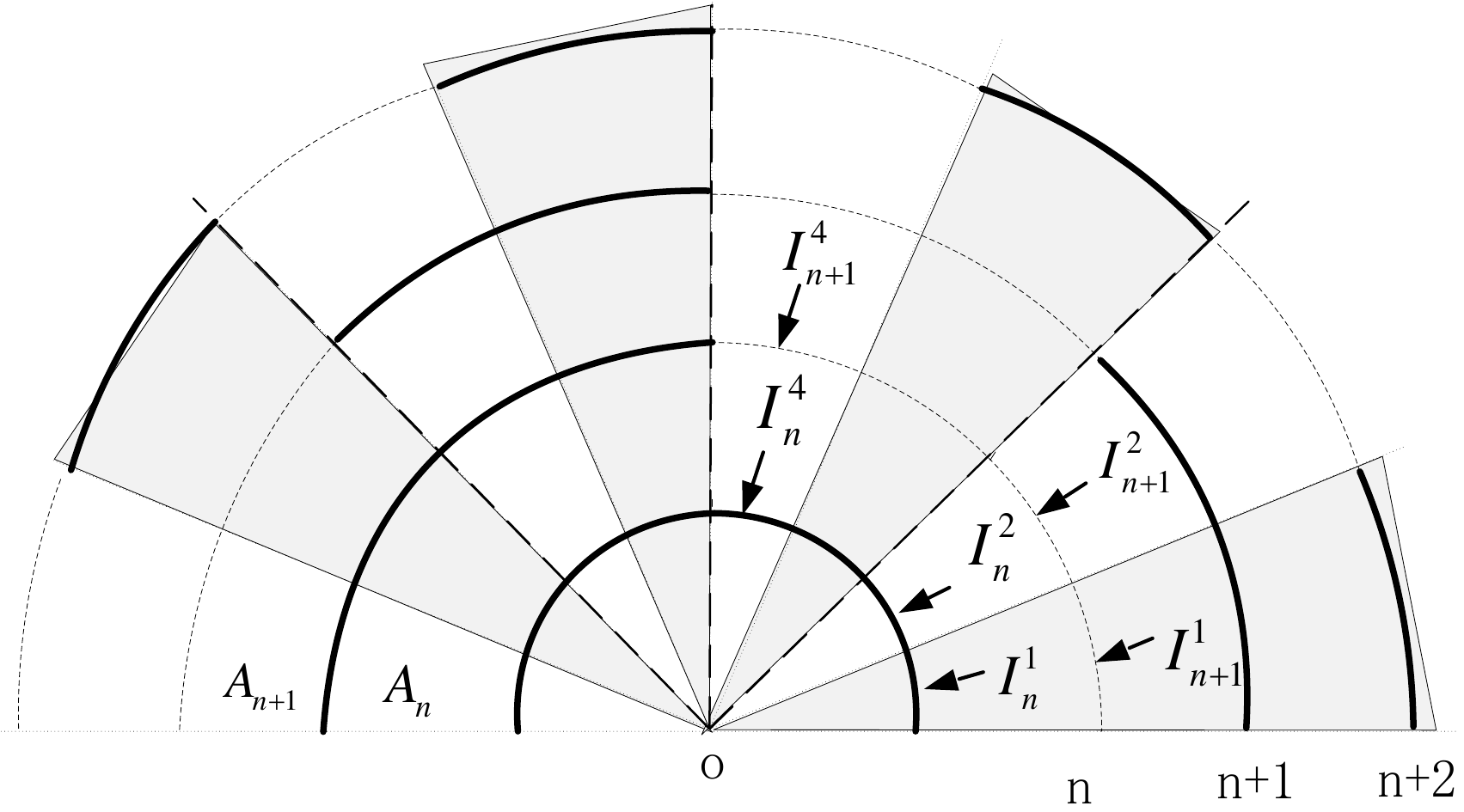}
	\captionsetup{singlelinecheck=off}
	\hfil
	\caption{An illustration of sectorized beam patterns in the cascade blockage model.}
	\label{fig5}
	\vspace{-0.25cm} 
\end{figure}
\subsection{Beamforming Model for UE}
{{The scenario retained for analyzing the case where
UEs have a beamforming capability
is based on the following assumptions:
\begin{itemize}
\item The UE is equipped with $2^k$ beams, with $k\ge 1$. 
\item The gain in the main lobe of a given beam is a constant $G \approx 2^k$ with beamwidth $2\pi/2^k$.
That is, the actual beam pattern is approximated by an ideal sector pattern.
as depicted by the shaded areas of Fig.\ref{fig5}.
On the uplink, all radiated power is hence
assumed to be concentrated in the main lobe, whereas
on the downlink, when activating beam $l$, then only the BSs located in this beam
interfere with the serving BS signal.
\item Whatever the beam, the serving BS signal on the downlink is still assumed to be provided by an
extra virtual BS with power 1 and
received with a Rayleigh fading.
\item The obstacle structure is the basic binary random model of the last
section with obstacles being independently present or absent with probability $p$ and $q$
respectively on the arcs $(\frac{2\pi}{2^n}l, \frac{2\pi}{2^n}(l+1))$, $l=1,\ldots,2^n$,
of the circle of radius $R_n$, for all $n\ge 1$. For the same reasons as above, we assume that $p\ge 1/2$ 
when considering the infinite obstacle model.
\item The UE beams are {\em in phase} with the obstacle structure.
Namely, the beam sectors are 
respectively on the arcs $(\frac{2\pi}{2^k}l, \frac{2\pi}{2^k}(l+1))$, $l=1,\ldots,2^k$.
\end{itemize}
}}
\noindent
Some of these assumptions are debatable. The last one for instance is rather specific and a random
phase of one structure w.r.t. the other would be more natural.
All these assumptions aim at making the mathematical model as tractable as possible.
The extension to less specific scenarios will be considered in subsequent papers.

We still focus on the downlink.
We adopt the Max SIR beam selection scenario. This means the UE calculates the SIR
in each beamforming direction and selects the beam with the maximal SIR.
{{The {\em best beam} within this setting is hence defined as
the beam which has the maximal SIR among all $2^k$ beams.}}

\subsection{Joint Distribution of Sector Interference}
\label{sec:fueq}
In this subsection $k$ is a fixed parameter. The UE is assumed
to be equipped with $2^k$ beam sectors.
Under the assumptions listed above, let $I_n^l$ denote the interference received by the UE in beam 
$l=1,\ldots,2^k$ and stemming from the complement of the closed ball of radius $R_{n-1}$.
Below, we first give iterative formulas allowing one to evaluate the joint Laplace transforms
\begin{align}
{\cal L}_n\left(s_1,\ldots,s_{2^k}\right):= \mathbb{E} \left[
\exp \left(-s_1I_n^1-\cdots-s_{2^k}I_n^{2^k}\right)\right],
\end{align}
for all $n\ge 0,$ and all $(s_1,\ldots,s_{2^k})$ in $\mathbb{R}_+^{2^k}$.
We then show that the coverage probability achieved by the
best SIR beam strategy (and other strategies as well) can be derived
from the knowledge of ${\cal L}_1(s_1,\ldots,s_{2^k})$.

We build a family of functions $H_{k},H_{k-1},\ldots,H_1$ by induction.
The function $H_{k}$ is the function of 1 real variable
\begin{align}
H_{k}(s):=\mathbb{E} \left[\exp \left(-sI_{k}^l\right)\right].
\end{align}
Note that, by symmetry, this function is the same for all $l$.

The function $H_{k-1}$ is the function of 2 real variables
\begin{align}
H_{k-1}(s,t):=
\mathbb{E} \left[ {\exp \left( { - sI_{k - 1}^{2j - 1} - tI_{k - 1}^{2j}} \right)} \right],
\end{align}
that is the joint Laplace transform of $(I_{k-1}^{2j-1}, I_{k-1}^{2j})$ at $(s,t)$.
This function is again the same for all $j, (1 \le j\le 2^{k-1})$. 

More generally, the function $H_{k-n}$ is the function of $2^{n}$ real variables
\begin{align}
	&  H_{k-n}(s_1,s_2,\ldots,s_{2^{n}})\notag\\
	&  :=\mathbb{E}\left[ e^{
		-s_{1}I_{k-n}^{2^{n}j-2^{n}+1}
		-s_{2}I_{k-n}^{2^{n}j-2^{n}+2}
		\cdots
		-s_{2^{n}-1}I_{k-n}^{2^{n}j-1}
		-s_{2^{n}} I_{k-n}^{2^{n}j} }\right].
\end{align}
By symmetry, this function is the same for all $j, (1 \le j\le 2^{k-n})$.
\begin{figure}[!t]
	\centering
	\includegraphics[width=2.8in]{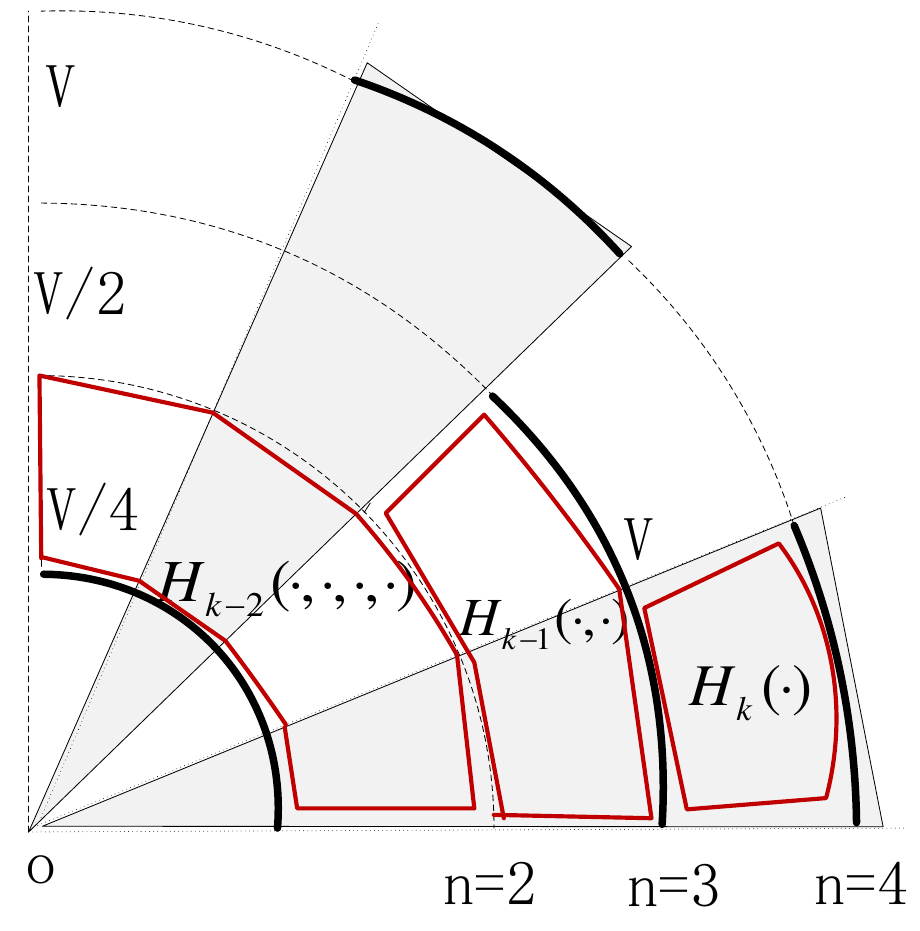}
	\captionsetup{singlelinecheck=off}
	\hfil
	\caption{A pictorial representation of the iteration procedure. The areas enclosed in red are the local boxes involved in $H_{k-n}(\cdot,...,\cdot)$.}
	\label{fig6}
	\vspace{-0.25cm} 
\end{figure}
The independence properties imply that
\begin{align}
	{\cal L}_{k}\left(s_1,\ldots,s_{2^k}\right)  &=  \prod_{l=1}^{2^k} 
	H_{k}(s_l),\notag\\
	{\cal L}_{k-1}(s_1,\ldots,s_{2^k})  &=  \prod_{j=1}^{2^{k-1}} 
	H_{k-1}(s_{2j-1},s_{2j}),
\end{align}
and more generally
\begin{align}
	& {\cal L}_{k-n}(s_1,\ldots,s_{2^k})\notag\\
	&  = \prod_{j=1}^{2^{k-n}} 
	H_{k-n}(
	s_{2^{n}j-2^{n}+1},
	s_{2^{n}j-2^{n}+2},\ldots,
	s_{2^{n}j-1},
	s_{2^{n}j}).
\end{align}
In particular
\begin{align}
	{\cal L}_{1}(s_1,\ldots,s_{2^k})=  
	H_{1}(s_{1},\ldots, s_{2^{k-1}})
	H_{1}(s_{2^{k-1}+1},\ldots, s_{2^{k}}).
\end{align}

By the same arguments as in the last section,
the function $H_{k}(s)$ is the solution of the functional equation
\begin{align}
H_{k}(s)=  \left\{pH_{k}^2(Ks)+qH_{k}^2(s)\right\} {\cal A}(s,V),
\end{align}
which is the analogue of (\ref{Lap}).
Similarly, $H_{k-1}(s,t)$ is obtained from $H_{k}$ through (see Fig.\ref{fig6})
\begin{align}
	&H_{k-1}(s,t) = \notag\\ 
	&\left\{ p H_{k}(Ks)H_{k}(Kt) + q H_{k}(s)H_{k}(t) \right\}{\cal A}(s,V/2){\cal A}(t,V/2)  .
\end{align}

More generally, if we know $H_{k-n+1}$
for some $1\le n< k$, then $H_{k-n}$ is obtained from $H_{k-n+1}$ through

\begin{align}
	& {H_{k - n}}({t_1}, \ldots ,{t_{{2^n}}})= \notag\\ 
	& \left\{ p {H_{k - n + 1}}(K{t_1}, \ldots ,K{t_{{2^{n - 1}}}})
	{H_{k - n + 1}}(K{t_{{2^{n - 1}} + 1}}, \ldots ,K{t_{{2^n}}})\right.\notag\\
	&  \left. + q
	{H_{k - n + 1}}({t_1}, \ldots ,{t_{{2^{n - 1}}}}){H_{k - n + 1}}({t_{{2^{n - 1}} + 1}}, \ldots ,{t_{{2^n}}}) \right\}\times\notag\\
	&   {\cal A}({t_1},V/{2^n}) \cdots {\cal A}({t_{{2^n}}},V/{2^n})\label{Mn}.
\end{align}

Note that $H_{k-n}$ is a function of $2^n$ variables.
So the memory requirement of the iterative associated with (\ref{Mn}) is proportional to $2^k$.
The numbers of function calls required to evaluate in $H_{k-n}(\cdot,...,\cdot)$ is also $2^n$
(because of the product ${\cal A}({t_1},V/{2^n}) \cdots {\cal A}({t_{{2^n}}},V/{2^n})$).
Hence the number of operations is also proportional to $2^k$, the number of beams.

\subsection{Coverage Probability under Best Beam Selection}
Under our assumptions, when denoting by $h^l$ the Rayleigh fading w.r.t. the BS signal in beam $l$
and by $G$ the {{directional}} gain,
the downlink $\theta$-coverage probability of the UE by the best beam is
\begin{align}\label{bestb}
	&P_{{\rm{cov}}}^{\max }(\theta )= 1 - {\mathbb P}\left[ \mathop {\max }\limits_l \left( \frac{Gh^l}{I_1^l}\right)  < \theta \right]  \notag\\
	&= 1 - {\mathbb P}\left[ \frac{{{Gh^1}}}{{I_1^1}} < \theta ,\frac{{{Gh^2}}}{{I_1^2}} < \theta ,...,\frac{{{Gh^{2^k}}}}{{I_1^{2^k}}} < \theta \right] \notag\\
	& = 1 - {\left( {{\mathbb E}\prod\nolimits_{l = 1}^{2^k}
			{(1 - {e^{ - \theta I_1^l/G}})} } \right)} \notag\\
	& = -\sum_{j=1}^{2^k} (-1)^j \notag\\
	&\sum_{1\le {i_1} < {i_2} <\cdots< {i_j}\le 2^k}
	\widetilde {\cal L}_1(s_{i_1},s_{i_2},\ldots,s_{i_j}){|_{{s_{i_1}} = \theta/G,...,{s_{i_j}} = \theta/G }} ,
\end{align}
where $\widetilde {\cal L}_1(s_{i_1},s_{i_2},\ldots,s_{i_j})$ stands for ${\cal L}_1$ evaluated at
the point $s=(s_1,\ldots,s_{2^k})$ with $l$-th coordinate equal to
$s_{i_m}$ for $l=i_m$ for $m=1,\ldots,j$, and 0 elsewhere. 

This shows that, as announced, the knowledge of ${\cal L}_1$ allows one to evaluate
the probability of coverage under the best beam selection strategy.
\begin{figure}[!t]
\centering
\includegraphics[width=2.8in]{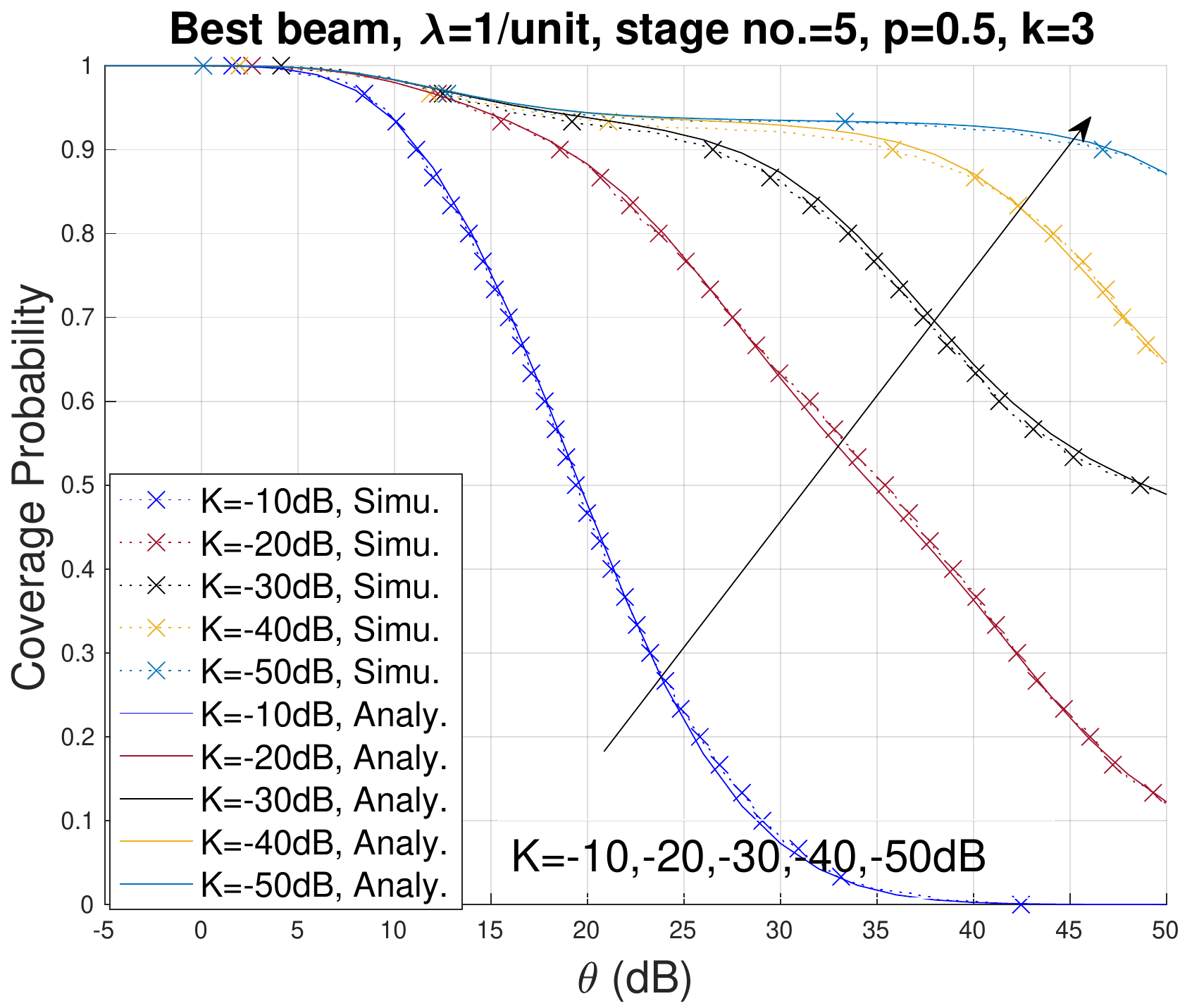}
\captionsetup{singlelinecheck=off}
\hfil
\caption{Best beam selection for different penetration losses $K$
($\lambda=1/unit$, max. stage no. $N=5$, $p=0.5$).}
	\label{fig7}
	\vspace{-0.25cm} 
\end{figure}

\subsection{Random Beam Selection Strategy}
As a baseline, we also consider a random beam selection strategy
where a beam is selected randomly as the serving beam for the typical UE.
Since the joint Laplace transform of the interference seen by the origin in the $2^k$ beams 
is ${\cal L}_1(s_1,\ldots,s_{2^k})$ with ${\cal L}_1$ the function defined in Subsection \ref{sec:fueq},
the Laplace transform of the interference in any beam is ${\cal L}_1(s,0,\ldots,0)$.  Hence the
coverage probability for SIR threshold $\theta$ is
$P_{{\rm{cov}}}^{\rm{rand} }(\theta )={\cal L}_{1}(\frac{\theta}{G},0,\ldots,0)$.

\subsection{Simulation Results}

\begin{figure}[!t]
	\centering
	\includegraphics[width=2.8in]{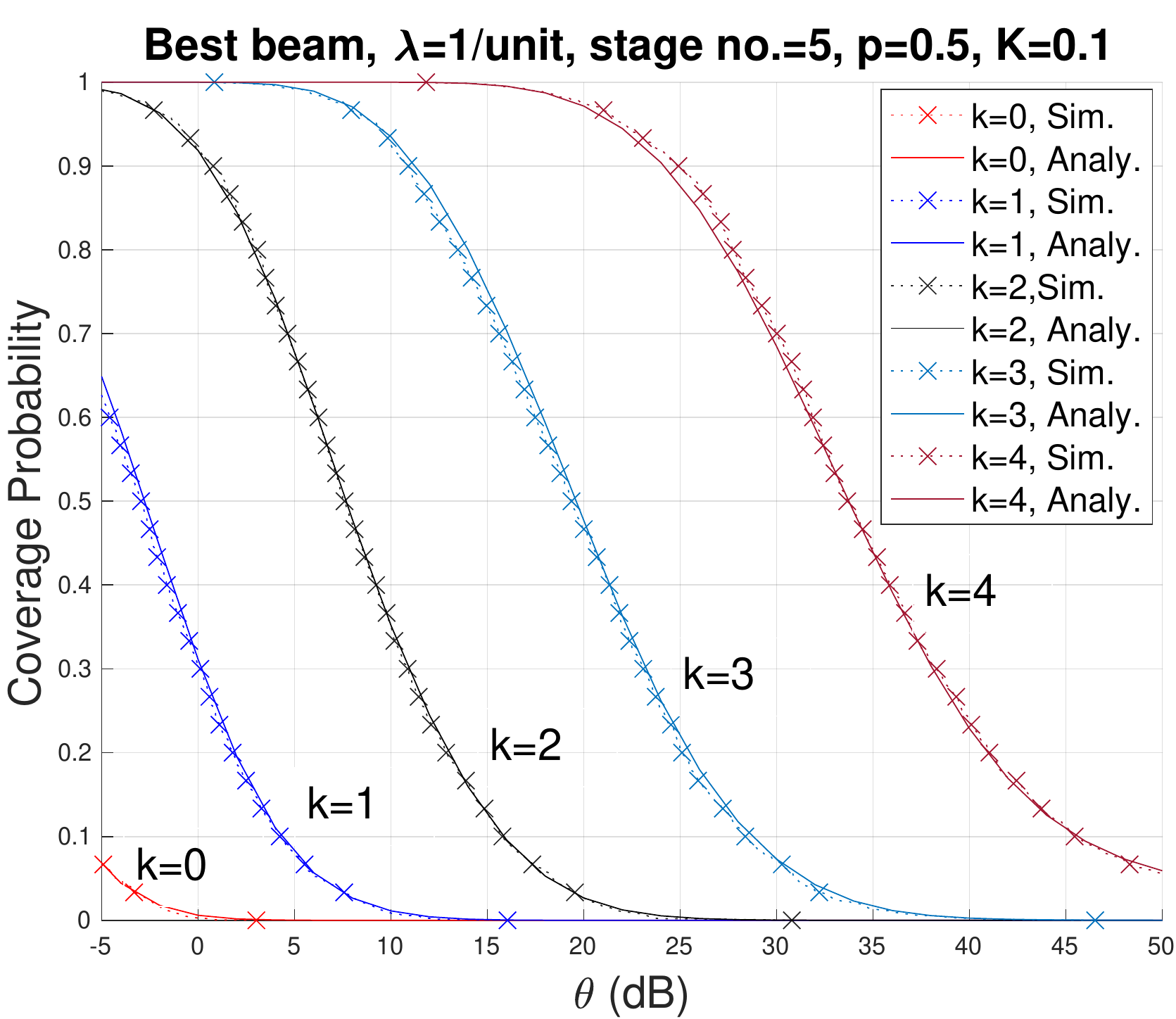}
	\captionsetup{singlelinecheck=off}
	\hfil
	\caption{Best beam coverage performance for different beam numbers. ($\lambda=1/unit$, max. stage no. $N=5$, $ p=0.5, K=0.1$)}
	\label{fig8}
	\vspace{-0.25cm} 
\end{figure}
\begin{figure}[!t]
	\centering
	\includegraphics[width=2.8in]{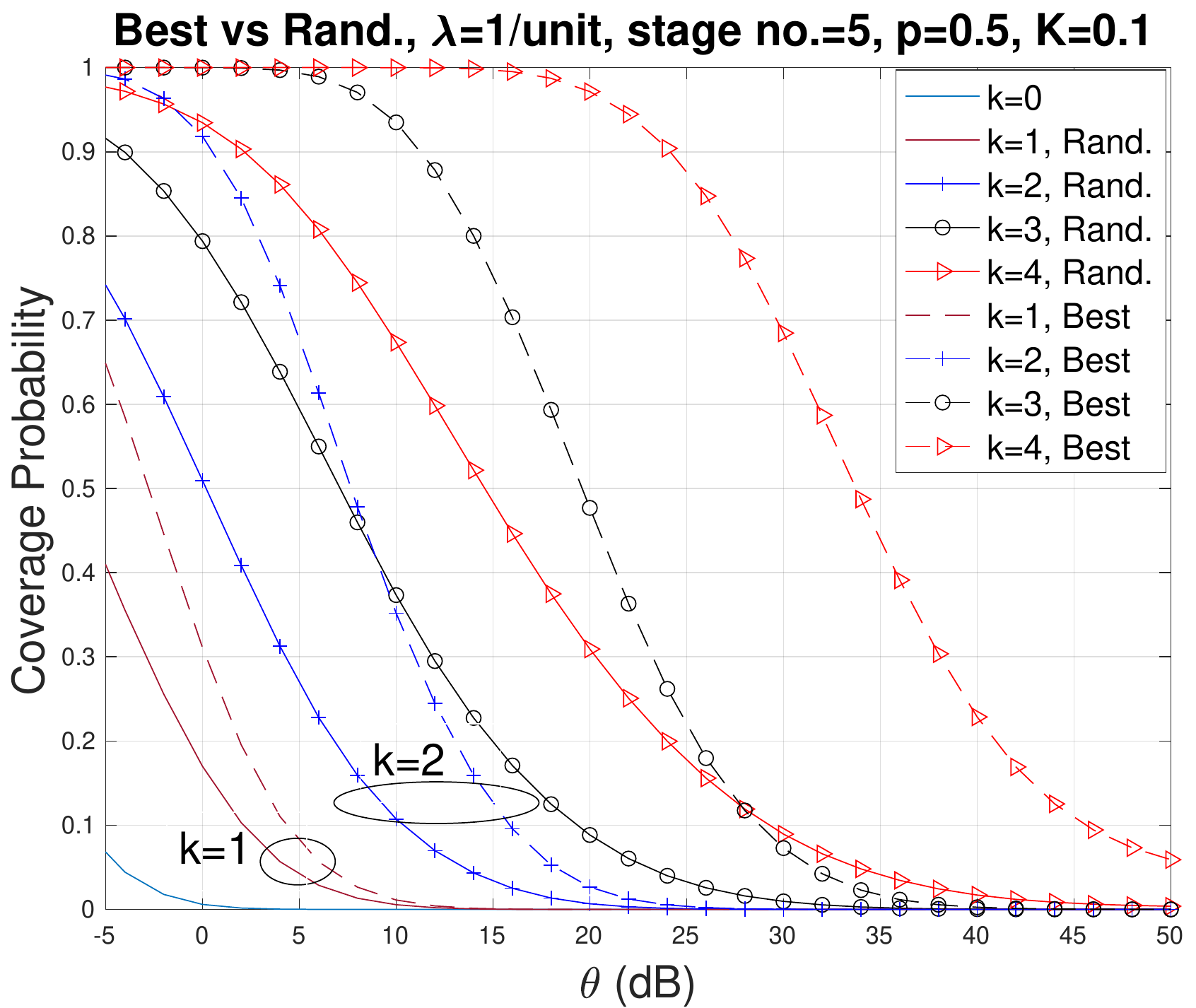}
	\captionsetup{singlelinecheck=off}
	\hfil
	\caption{A comparison between best beam and random beam selections. ($\lambda=1/unit$, max. stage no. $N=5$, $ p=0.5, K=0.1$)}
	\label{fig9}
	\vspace{-0.25cm} 
\end{figure}

Fig.\ref{fig7} compares the coverage probability of the best beam scheme for different penetration losses
when $N=5$, $\lambda=1$. Our analytical results match well with simulation. As expected, when
penetration loss increases, the coverage probability increases due to stronger blockage effects. 

The gain of best beam selection are shown in Fig.\ref{fig8} The log in base 2 of the number of beam number varies
from $k=0$ to $k=4$, and $K$ is set to 0.1. When $k=0$, the UE is almost uncovered due to strong LOS
interference. When $k$ increases, the coverage probability increases dramatically thanks to 
higher beam selection diversity, higher directional gain and narrower beamwidth.
This in a sense justifies the use beamforming in 5G networks to compensate path loss at mmWave frequencies.  

Interestingly, when the number of beams doubles, the gain obtained by the best beam selection
over a random beam selection increases as shown in Fig.\ref{fig9}. When $k$ is small, this performance 
gain is not significant, particularly so at high thresholds, e.g., almost no gain for $k=1$, and a
$25\%$ gain for $k=2$, when $\theta=10dB$. This means when $k$ is small, a random beam selection
does almost as well as best beam selection, at least when the SIR target is high.
In contrast, when $k$ is large, the beam scanning and selection procedure are justified.

\section{Beam Switching in UE}
In this section, we study beam switching.
For a static UE equipped with multiple antennas, beam switch could happen for, e.g., the following reasons:
\begin{itemize}
\item Deep fade: the current UE beam towards the serving BS suffers from a deep fade,
which could not be corrected by conventional physical layer techniques, such as channel coding,
interleaving, antenna diversity etc. 
\item Mobile blockage:  the blockages caused by mobile blockers (e.g., a vehicle)
could significantly impair the received signal strength and system performance. 
\item UE rotation:  for hand-held devices, the UE is rotated and the current beam is
obstructed due to a different holding gesture.
\end{itemize}
All the above scenarios demand prompt beam switch procedures in order to reduce the outage duration.
Since any exhaustive beam sweeping comes with a large delay of synchronization, signal detection,
and reference signal quality evaluation for each beam pair, fast association and beam switch
procedures are desirable. 

A simple procedure is that where the UE operates a switch to a new beam with a given
angular separation from the original beam.
In this context, it is interesting to investigate the conditional probability
that the UE is covered in a target beam conditional on the fact that it is covered {{(or not covered)
in the initial (or source) beam}}.
Without loss of generality, we assume the source beam is the first beam and
the target beam is that with index $l$. Taking $l$ close to 1 (mod $2^k$) means a small
angular switch, whereas taking it far away from $1$ means a large angular switch. 

{{The conditional coverage probability can be obtained from the joint Laplace transform discussed in 
Section \ref{sec:fueq} through the relation}}
\begin{align}
	P_{{\mathop{\rm cov}} }(\theta,l|\theta,1 ) &: =\frac{ \mathbb{P}\{ SI{R_1} > \theta ,SI{R_l} > \theta \}}{\mathbb{P}\{ SI{R_1} > \theta \}}\notag\\ 
	&=\frac{{\cal L}_{1}(\frac{\theta}G, 0,\ldots,0,\frac{\theta}{G}, 0,\ldots,0)}{{\cal L}_{1}(\frac{\theta}{G} ,0,\ldots,0) },
\end{align}
where in the last numerator, the second non-zero argument is for variable $s_l$.
{{Other conditional probabilities (like the chance to be covered in the target beam given
there is no coverage in the source beam) can easily be deduced from this.}}

\begin{figure}[!t]
	\centering
	\includegraphics[width=2.8in]{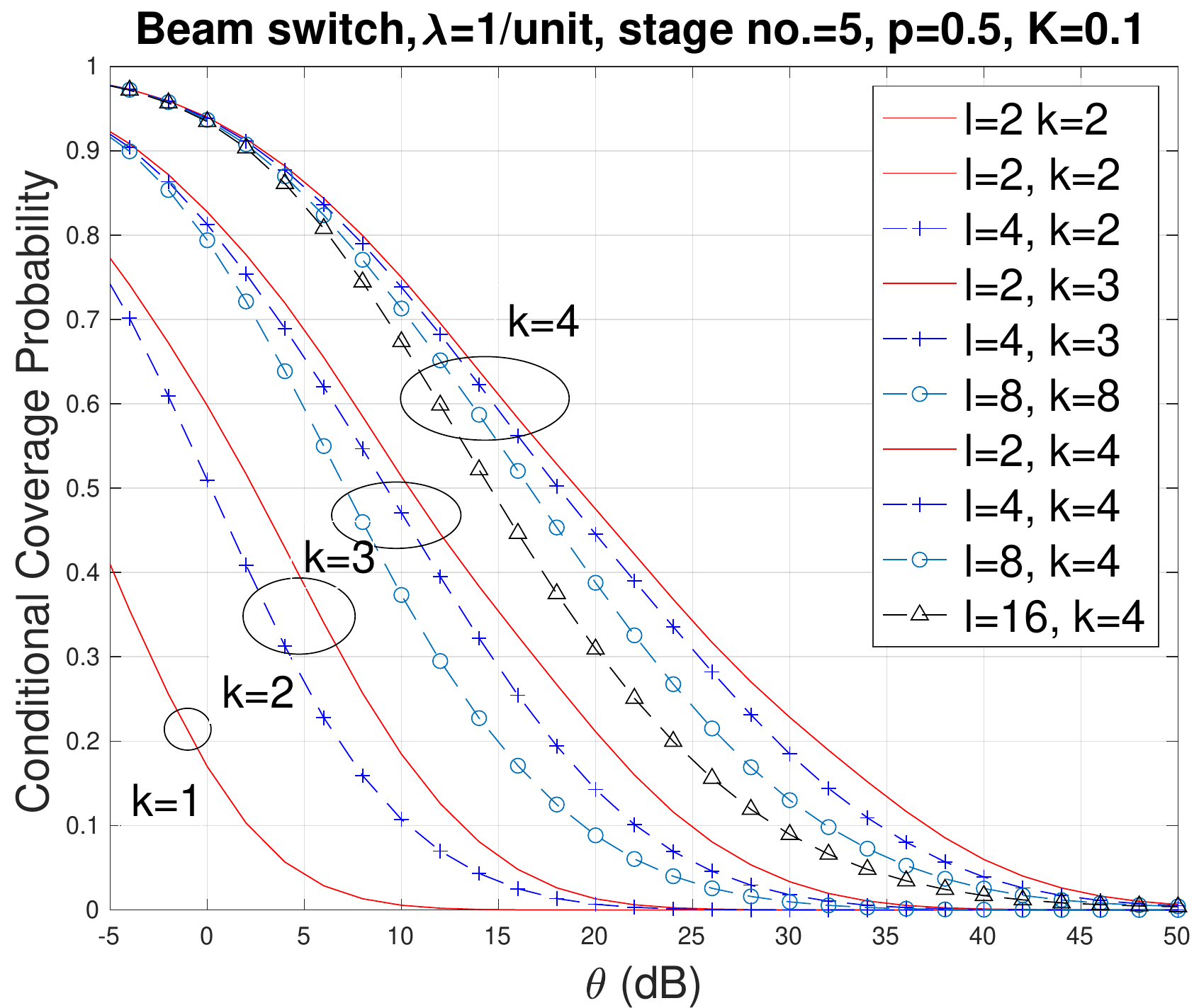}
	\captionsetup{singlelinecheck=off}
	\hfil
	\caption{A comparison of conditional coverage probability of beam switch among different beam patterns. ($\lambda=1/unit$, max. stage no. $N=5$, $p=0.5, K=0.1$)}
	\label{fig10}
	\vspace{-0.0cm} 
\end{figure}

\subsection{Simulation Results}
In Fig.\ref{fig10}, for each beam pattern, when the beam index $l$ increases, the conditional coverage probability degrades as the target beam has more independent blockages. For example, for 16-beam pattern, the 2-th beam has the largest conditional coverage performance because it shares the maximal common blockages with the source one (i.e., the 1-th beam). When beam index increases to 4, only common blockages of the first and second stages are shared by source and target beams. Hence, conditional coverage probability decreases. When beam index varies from 5 to 8, only first stage blockage is shared by them. When index varies from 9 to 16, target beams have complete independent blockage environments and performance metric achieves the minimum.
\subsection{Beam Switch Strategy Modification}
The above analysis motivates us to modify the classical beam switch strategy for static UEs. In the modified beam switch procedure, the static UE alternatively sweeps beams around the source one, and sequentially increase beam index until a satisfied beam is obtained, instead of sweeping beam towards uni-direction until whole beam space is scanned. This modification takes the advantage of the fact that in the correlated blockage/shadowing environment, the adjacent BSs are often correlated over short space distance/angle \cite{ref1}. In other words, if a BS has satisfied coverage performance before switch, its nearby BSs are likely to have similar performance,  as shown in Fig.\ref{fig10}. This fast switch strategy can shorten association duration during beam space scanning.  

\section{Conclusion}
This paper proposes a multiplicative cascade blockage model to emulate correlated blockage environments.
This model is a complement to the Manhattan-type urban model and random blockage models.
This model leads to new iterative algorithms for the Laplace transform of interference
for omni-directional UE. 
Another iterative algorithm is derived to analyze the coverage probability
under the best beam selection for beamforming capable UEs.
A further analysis of conditional coverage probability shows the benefit of correlation on beam switch.
It is also shown that: \emph{1)}  Sparse blockage environments have inferior coverage
performance compared to dense blockage environments. \emph{2)} Blockage correlation effects
can improve coverage performance in comparison with independent blockage environments.
\emph{3)} The best beam selection algorithm is very effective for beamforming UEs
to compensate path or blockage penetration loss.
\emph{5)} Correlation brought by blockages can be leveraged to accelerate beam swtich and association procedures.
{This paper is the first attempt to use multiplicative cascades in blockage effect problem and Rayleigh fading assumption is made for computational tractability. In the future work, other fading models, such as Rician fading or Nakagami fading will be incorporated into this cascade blockage model for more general cases.}
\section{Acknowledgements}
The present authors would like to thank Prof. Philippe Martins for suggestions and discussions in preparing this paper.

\ifCLASSOPTIONcaptionsoff
\newpage
\fi

\end{document}